\documentclass{appolb}
\usepackage{epsfig}

\begin{document}

\title{Exclusive Dijet production from CDF2LHC
\thanks{Presented at the ``TeV4 LHC'' Workshop, February 3-5, 2005, Brookhaven NY, USA.}
}
\author{Michele Gallinaro \footnote{Representing the CDF collaboration.} 
\address{The Rockefeller University \\ 1230 York Avenue, New York, NY 10021, USA}
}
\maketitle
\begin{abstract}

Exclusive dijet production at the Tevatron can be used as a benchmark to establish predictions on exclusive diffractive Higgs production, 
a process with a much smaller cross section.
Exclusive dijet production in Double Pomeron Exchange processes,
including diffractive Higgs production with measurements at the Tevatron and predictions for the Large Hadron Collider are presented.
Using new data from the Tevatron and dedicated diffractive triggers, no excess over a smooth falling distribution for exclusive dijet events could be found.
Upper limits on the exclusive dijet production cross section are presented and compared to current theoretical predictions.

\end{abstract}

\section{Introduction}

The search for the Higgs boson occupies the center-stage of the high-energy physics program, both
currently at the Tevatron and in the near future at the Large Hadron Collider (LHC) at CERN. 

Within the standard model (SM)~\cite{sm}, the Higgs mechanism invoked to break the electroweak symmetry
implies the existence of a single neutral scalar particle, the Higgs boson. 
The mass of this particle is not specified, but indirect experimental limits are 
obtained from precision measurements of the electroweak parameters.
Currently, these measurements 
constrain its value to less than 260~GeV/c$^2$ at 95\% confidence level~\cite{higgs_limit}.
Indications that a few Higgs candidate events were found at $M_H\simeq 115~{\rm GeV/c^2}$ during the last phase of data-taking 
at the large electron-positron (LEP) collider at CERN attracted world-wide 
attention in the scientific community.
Findings were later dismissed and a lower Higgs mass limit at $M_H>114.4~{\rm GeV/c^2}$ (95\% C.L.) was set~\cite{sm_higgs}, 
but interest has remained high.

\begin{figure}[h]
\epsfxsize=1.0\textwidth
\centerline{\epsfig{figure=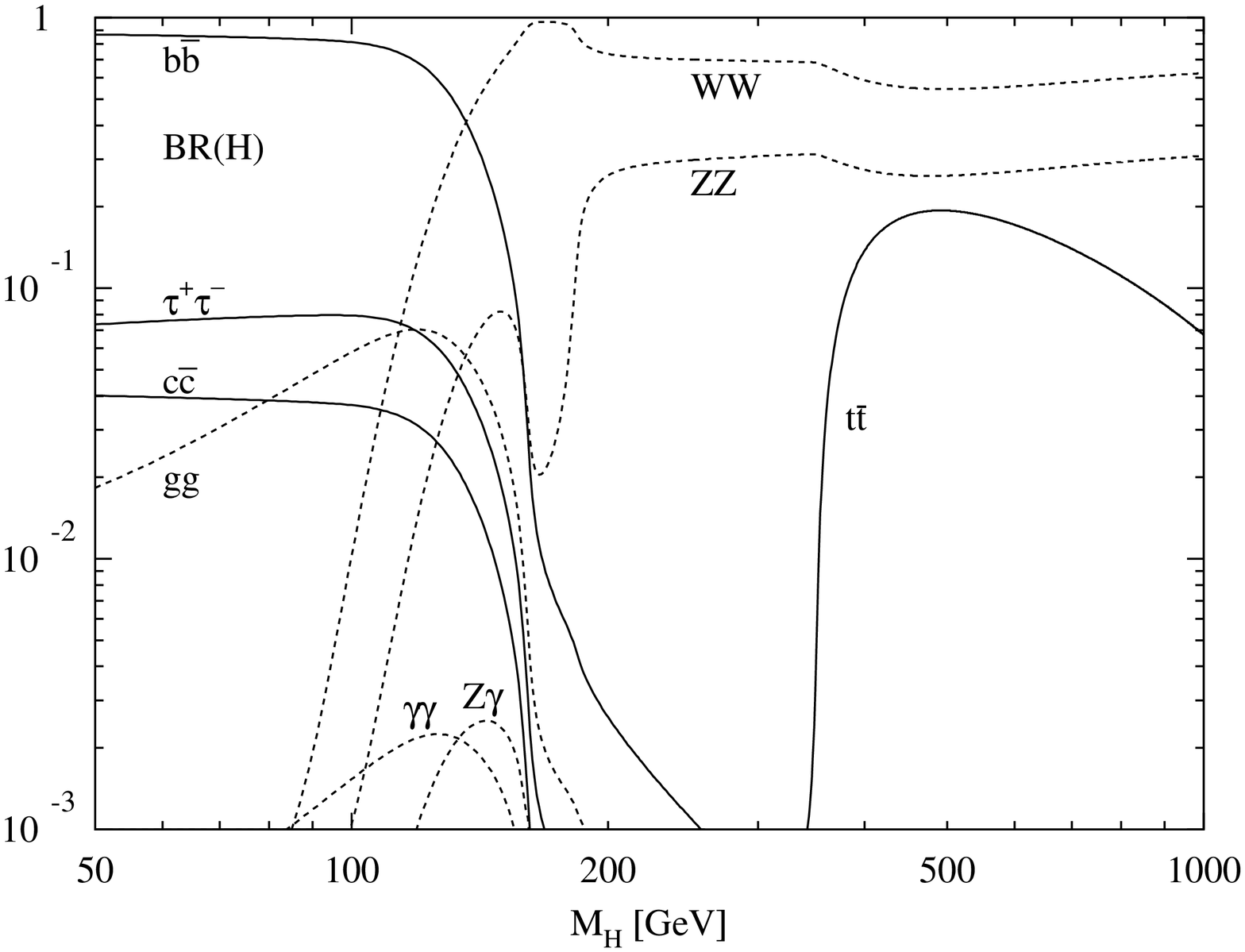,width=0.69\hsize}}
\caption{\label{fig:higgs_dec}
\small
Branching ratios of the standard model Higgs decay channels for different Higgs masses (from Djouadi et al., 1998~\cite{higgs_decays}).
}
\end{figure}

In the case of a small mass ($M_H<130~{\rm GeV/c^2}$),
the Higgs boson decays predominantly to $b\bar b$ or $\tau^+\tau^-$ pairs, 
with branching fractions of $\sim 90\%$ and $\sim 10\%$, respectively (Fig.~\ref{fig:higgs_dec}). 
In this case, the identification of Higgs production and decay modes will be complicated by the large backgrounds~\cite{higgs_prod}.
Diffractive processes with lower branching ratios
may result in a cleaner experimental signature and are worth exploring.
This is the case in Higgs production through Double Pomeron Exchange (DPE) processes
$$ p\bar p \rightarrow p H \bar p  \hspace*{0.8cm} {\rm or }  \hspace*{0.8cm} p p \rightarrow p H p , $$
where the leading hadrons in the final state are produced at small angles with respect to the direction of the incoming particles
and two large forward rapidity gap regions are present on opposite sides of the interaction.
The exclusive Higgs process was originally studied by Bialas and Landshoff~\cite{bialas}, 
and then followed by other theoretical work, such as that by Khoze, Martin and Ryskin~\cite{khoze}.
Diffractive Higgs production could provide a distinct signature with  
exclusive two-jet ($b\bar b$ or $\tau^+\tau^-$) event final states.
Furthermore, the presence of the rapidity gaps provides an experimental environment which is practically
free of background containing soft secondary particles, and where the signal to background event ratio is favorable. 
The background from direct $b\bar b$ production is small thanks to several suppression mechanisms
(such as color and spin factors, and the $J_z=0$ selection rule)~\cite{khoze_chi}.
Therefore, the signal from $H\rightarrow b\bar b$ is expected to have a mass resolution which is
greatly improved due to the absence of underlying event particles.

The predictions for the Higgs cross section due to exclusive DPE production are model dependent.
In one model~\cite{khoze_higgs},
which at the time of writing has still survived the exclusion limits set by the Tevatron data, and for a 
Higgs mass of $M_H=120~{\rm GeV/c^2}$, 
the predicted cross sections are $\sigma_H^{\rm TeV}\sim 0.2$~fb at the Tevatron 
and $\sigma_H^{\rm LHC}\sim3$~fb at the LHC, with large uncertainties. 
In such a calculation, a signal over background ratio of $S/B\sim 4$
is expected when a mass resolution of $\sim 1$~GeV and a $b$-jet fake rate of 1\% are taken into account.
When a more realistic estimate for the mass energy resolution ($\sim 3$~GeV) and the signal efficiency 
($\sim 6\%$) are considered~\cite{boonekamp},
the signal over background ratio decreases, thus requiring more data to extract a Higgs signal.
Since only a handful of events are expected for each $100 ~{\rm fb^{-1}}$ of data at the LHC, 
this channel may be hard to unveil.

\section{Exclusive Dijet Production}

The exclusive dijet production rate in DPE events, discussed in the previous section, is of great interest in determining
the (background to) exclusive Higgs production cross section and in preparation for the LHC experiments~\cite{higgs}.
The gluon-gluon fusion Higgs production process $g g\rightarrow H$ is replaced by the 
$gg\rightarrow {\rm jet~ jet}$ process, with a much larger production cross section.
Therefore, measurements at the Tevatron can directly provide a background estimate, and a benchmark 
for predicting the exclusive Higgs production cross section.
The characteristic signature of this type of events is a leading nucleon and/or a rapidity gap on both 
forward regions, and it results in an exclusive dijet final state produced together
with both the leading proton and anti-proton surviving the interaction and escaping in the very forward region.
At CDF, the Roman Pot (RP) spectrometer~\cite{fd} can tag the anti-proton, while
the proton is inferred by the presence of an adjacent large ($\Delta\eta>3$) rapidity gap.

\begin{figure}[tp]
\epsfxsize=1.0\textwidth
\centerline{\epsfig{figure=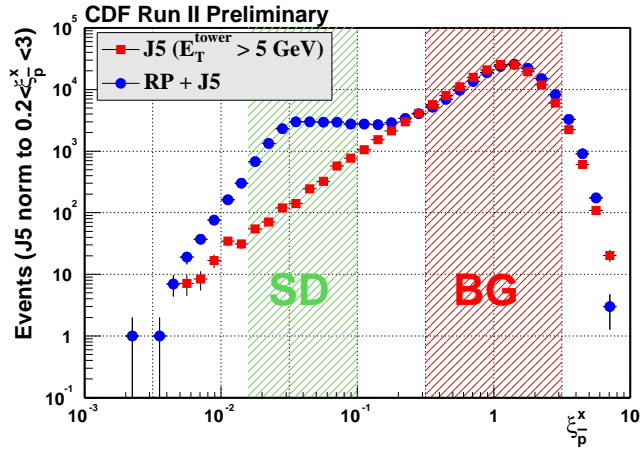,width=0.75\hsize}}
\caption{\label{Xi} 
Momentum loss of the antiproton ($\xi_{\overline{p}}$) distribution in the RP+J5 and J5 samples.
SD and BG regions are selected according to the measured $\xi$ values.
}
\end{figure}

During Run~I, about 100 DPE candidate events were identified and used to set an upper limit of 3.7~nb
on the exclusive dijet production cross section~\cite{dpe}.
In Run~II, a dedicated trigger (RP+J5) selects events with a three-fold RP coincidence and at least one
calorimeter tower with $E_T>5$~GeV~\cite{fd}.
A further offline selection requires at least two jets of $E_T^{corr}>5$~GeV and $|\eta|<2.5$.
Jets are corrected for detector effects and underlying event background contributions.
Calorimeter information alone is used to determine $\xi_{\overline{p}}=\frac{1}{\sqrt{s}}\sum_{i=1}^nE_T^ie^{-\eta^i}$, 
which is calculated using all calorimeter towers (Fig.~\ref{Xi}).
The declining of the distribution at $\xi_{\overline{p}}\sim 0.03$ occurs in the region where the RP acceptance is decreasing. 
The contribution in the large number of events at $\xi_{\overline{p}}\sim 1$ comes from two sources:
diffractive dijets with a superimposed soft non-diffractive interaction, and non-diffractive dijets superimposed with a soft diffractive interaction.
The plateau observed between $0.02<\xi_{\overline{p}}<0.1$ (SD) results from a $d\sigma/d\xi\sim 1/\xi$ distribution, which is expected for diffractive production.

\begin{figure}[tp]
\epsfxsize=1.0\textwidth
\centerline{
\epsfig{figure=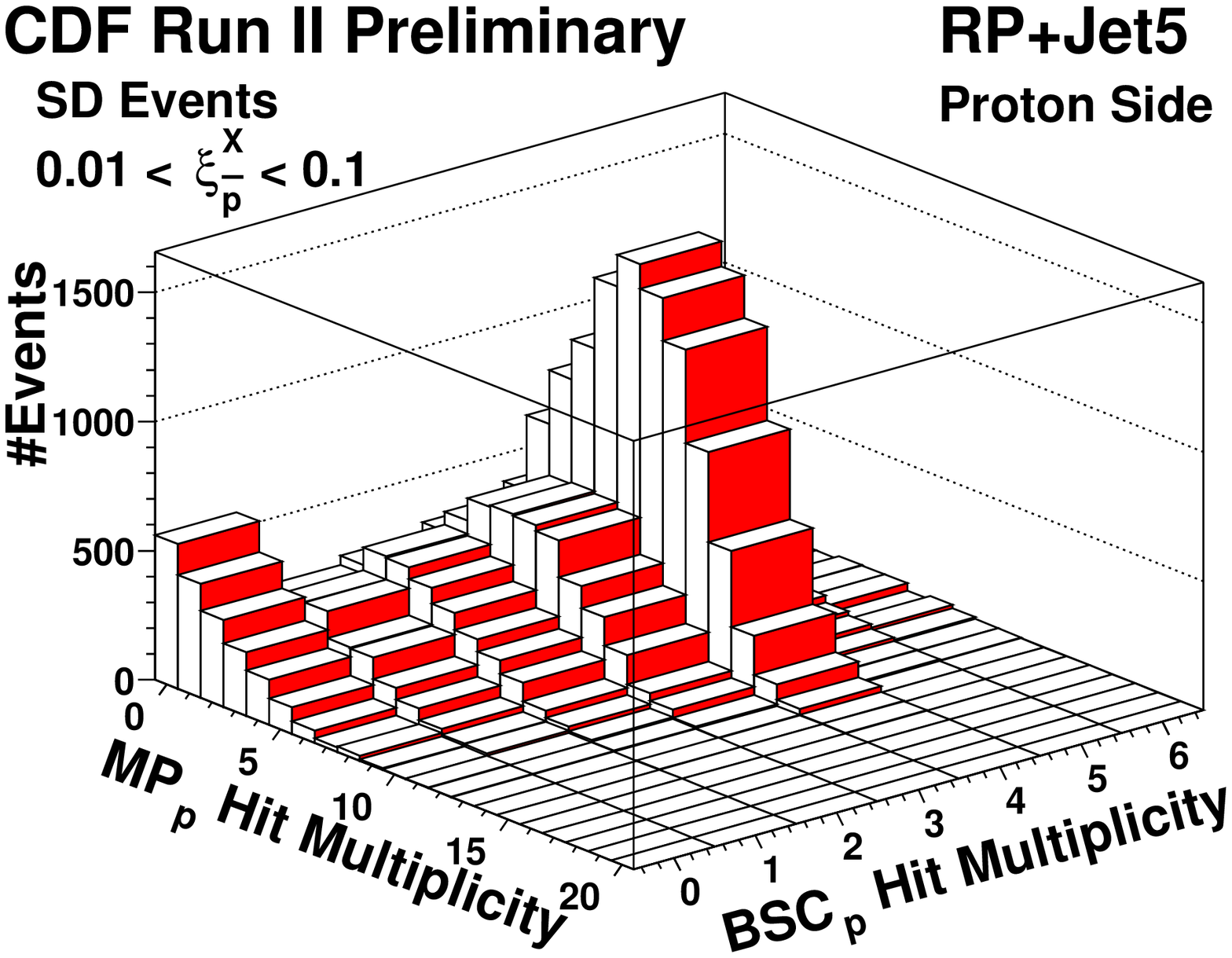,width=0.50\hsize}
\epsfig{figure=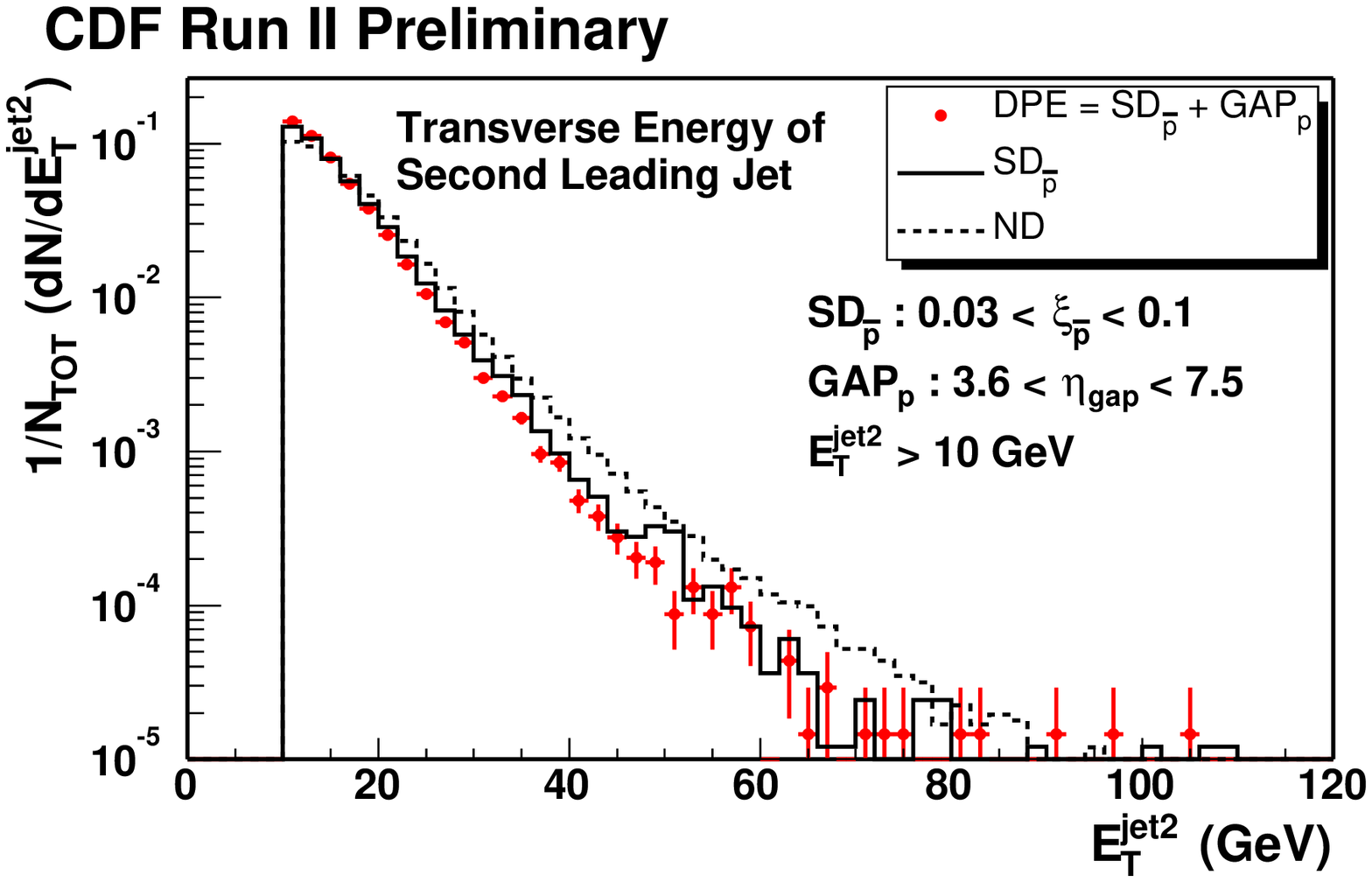,width=0.56\hsize}
}
\caption{\label{dpe_et_dphi}
{\em Left}: MP versus BSC multiplicity on the outgoing proton side in RP+J5 triggered events;
{\em Right}: Next-to-leading jet transverse energy distribution.
}
\end{figure}

The RP+J5 trigger can also be used to study DPE events, which can be isolated
by counting BSC (Beam Shower Counters) and MP (MiniPlug) multiplicities~\cite{bscmp} on the proton side (Fig.~\ref{dpe_et_dphi}, left). 
The two peaks, at high and low multiplicity, are due to Single Diffractive (SD) and DPE events, respectively.
A much larger sample has already been collected with a dedicated trigger
requiring one RP coincidence, a proton-side (opposite to the RP-side) rapidity gap in the BSCs, and at least one 
calorimeter tower ($E_T>$5~GeV) in the central detector.
Multiple interactions are rejected offline by requiring events with 0 or 1 vertices.
At least two jets ($E_T^{corr}>$10~GeV, $|\eta|<2.5$) are required.
The sample is further tightened by requiring the events to have $0.01<\xi_{\overline{p}}<0.1$. 
In order to reduce multiplicity fluctuations in SD events and thus enhance DPE events, 
a rapidity gap of $\sim 4$ units, including MP and BSC ($3.6<|\eta|<7.5$), is also required on the proton side.
The $E_T$ distributions for both leading and next-to-leading jets 
are similar for ND, SD and DPE samples (Fig.~\ref{dpe_et_dphi}, right).

\begin{figure}[h]
\epsfxsize=1.0\textwidth
\centerline{
\epsfig{figure=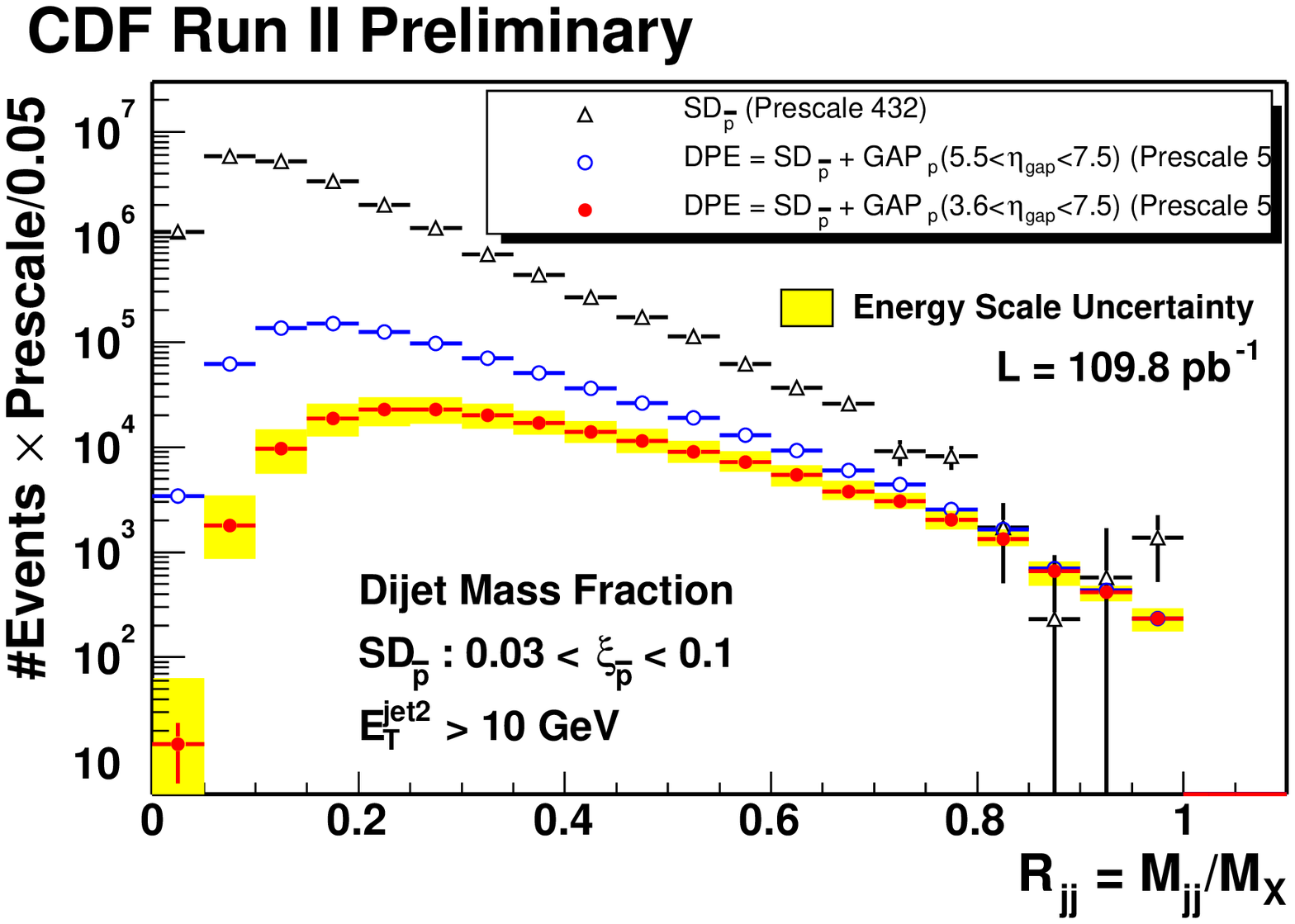,width=0.56\hsize}
\epsfig{figure=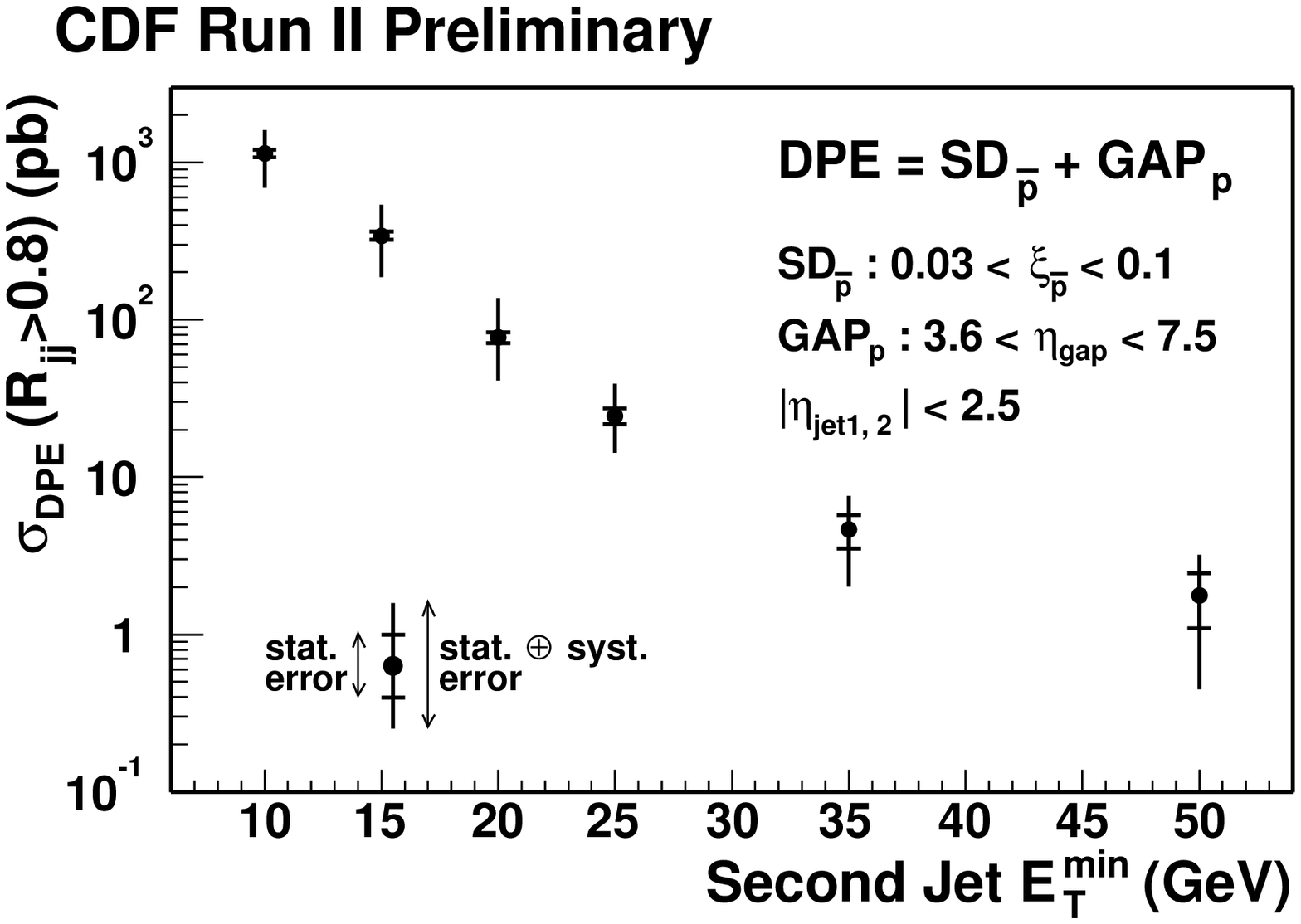,width=0.56\hsize}
}
\caption{\label{fig:dpe_massratio}
\small
{\em Left:} Dijet mass fraction for different rapidity gap regions.
{\em Right:} Upper limit cross section for DPE dijet events with $R_{jj}>0.8$ as a function of the minimum next-to-leading jet $E_T$.
}
\end{figure}

At CDF, in $\sim 110~{\rm pb^{-1}}$ of Run~II data, the DPE dijet sample already consists of approximately 17,000 events.
The dijet mass fraction ($R_{jj}$), defined as the dijet invariant mass ($M_{jj}$) divided by the mass of the entire system,
$M_X =\sqrt{\xi_{\overline{p}}\cdot\xi_p\cdot s}$, is calculated using all available energy in the calorimeter.
If jets are produced exclusively, $R_{jj}$ should be equal to one.
Uncorrected energies are used in Figure~\ref{fig:dpe_massratio} (left) and no visible excess is evident 
at $R_{jj}\sim 1$ over a smooth distribution.
After including systematic uncertainties, an upper limit on the exclusive dijet production cross section is calculated
based on all events with $R_{jj}>0.8$ (Table~\ref{tab:xs}). 
The measurement provides a generous upper limit cross section, as all events at $R_{jj}>0.8$ 
are considered due to exclusive dijet production.
The cross section upper limits on exclusive dijet production as a function of the minimum next-to-leading jet $E_T$ 
is presented in Figure~\ref{fig:dpe_massratio} (right).

\begin{table}[h]
\begin{center}
\begin{tabular}{c|c}
\hline
minimum leading jet $E_T$ & cross section upper limit \\ \hline
10 GeV & $1140\pm60({\rm stat})^{+47}_{-45}({\rm syst})$~pb \\
25 GeV & $ 34\pm3({\rm stat})^{+15}_{-10}({\rm syst})$~pb \\ \hline
\end{tabular}
\caption{\label{tab:xs} Exclusive dijet production cross section limit for events at $R_{jj}>0.8$.}
\end{center}
\end{table}

A brief note of clarification is due.
The measurement of the dijet mass fraction is only minimally affected by the uncertainty in the calorimeter energy scale.
In fact, the uncorrected energies cancel out the uncertainties in the ratio. Furthermore, at large dijet mass fractions, 
where the exclusive production is expected to emerge, only a small energy is found outside the two main energy clusters,
and therefore its uncertainty is small relative to the entire event reconstructed energy.
Thus, the upper limit, which is calculated for all events with $R_{jj}>0.8$, is indeed generous.

\begin{figure}[tp]
\epsfxsize=1.0\textwidth
\centerline{
\epsfig{figure=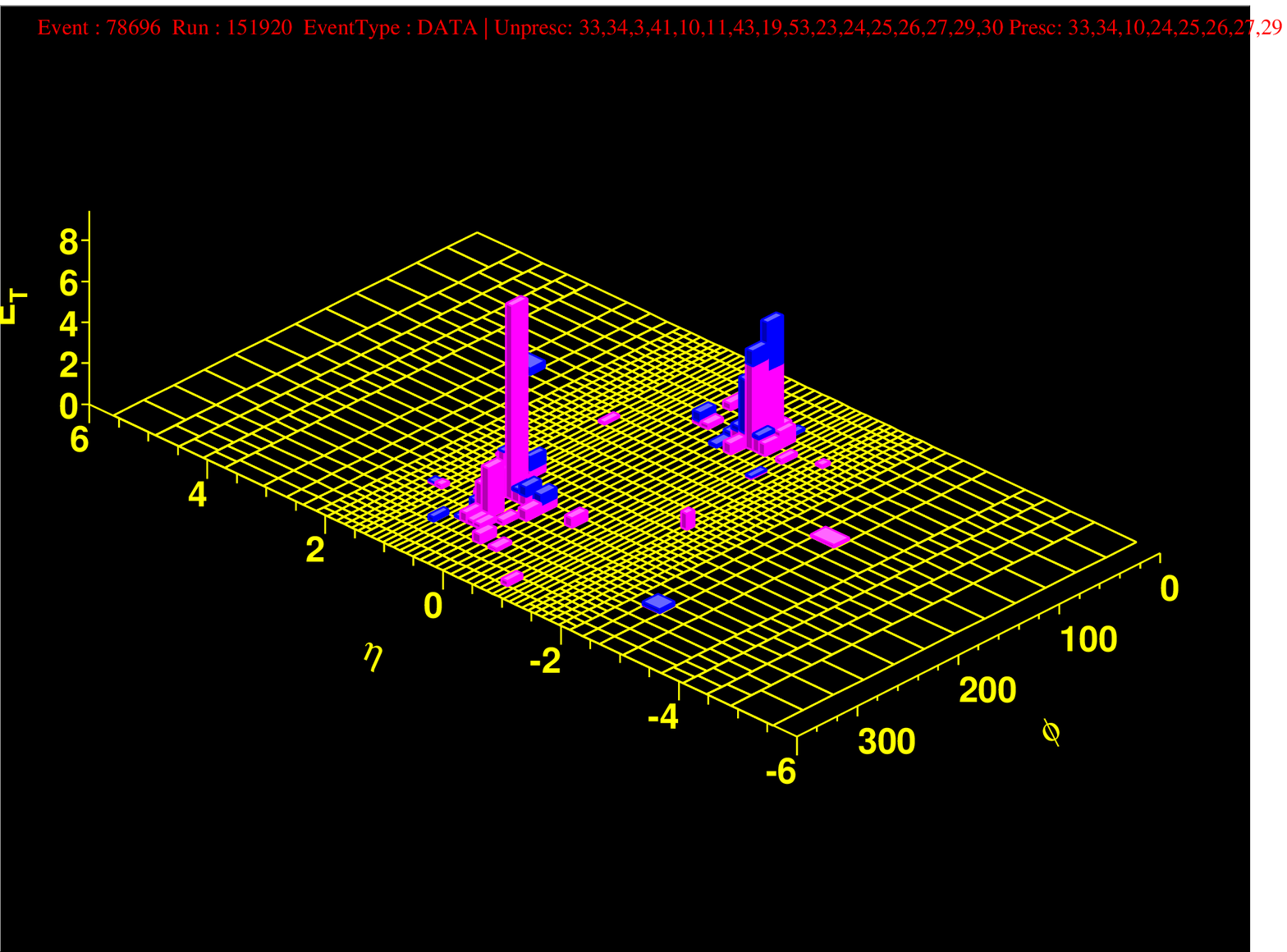,width=0.56\hsize}
\epsfig{figure=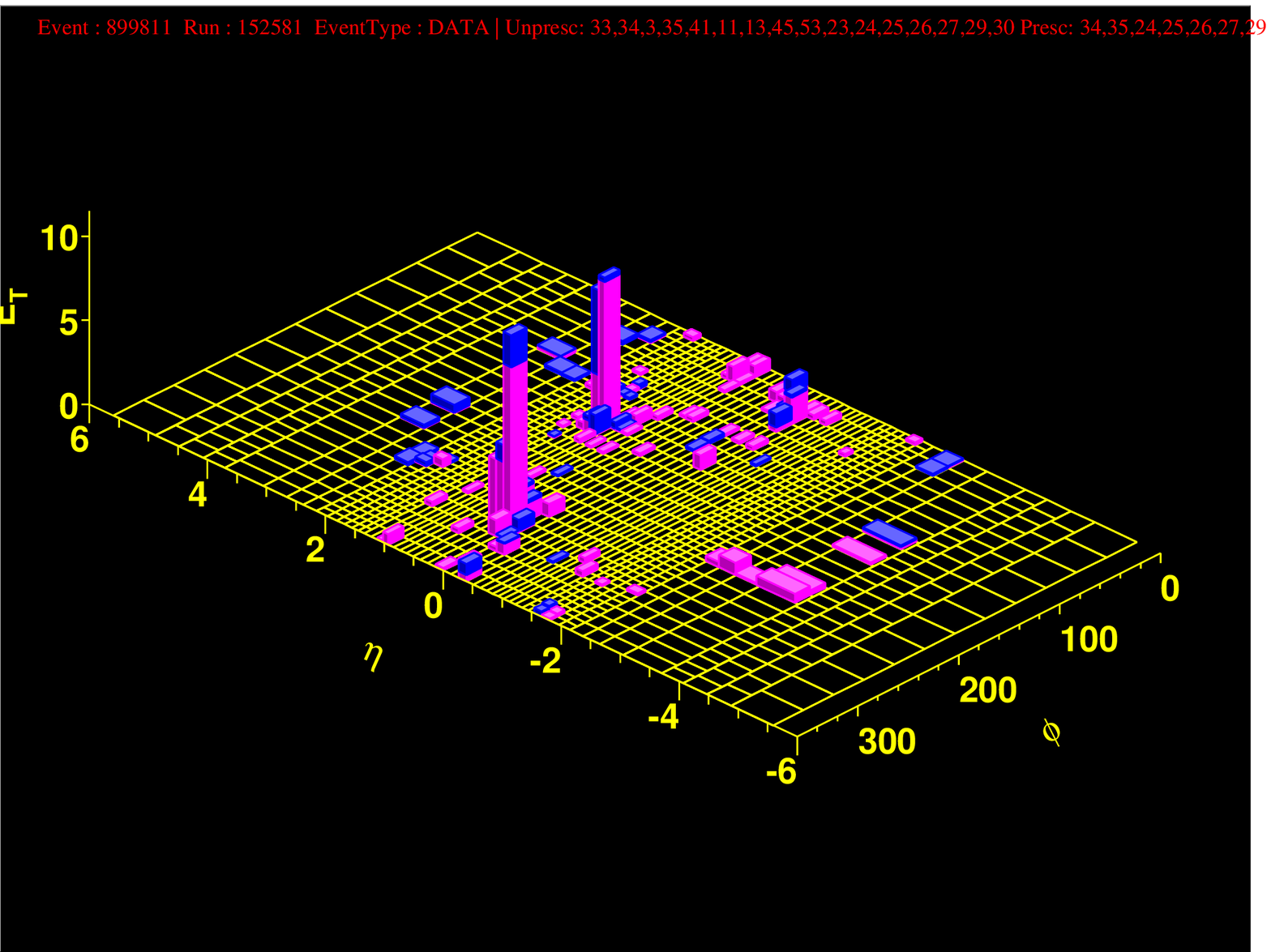,width=0.56\hsize}
}
\caption{\label{fig:event_display}
Event display in the pseudorapidity-azimuthal angle, $\eta-\phi$, coordinates of two dijet events with:
({\em left}) large mass fraction ($R_{jj}=0.81$) and
({\em right}) small mass fraction ($R_{jj}=0.36$).
The event with large mass fraction shows calorimeter energy deposition well outside the main thrust of both jets, suggesting the two jets are not produced exclusively.
}
\end{figure}

Figure~\ref{fig:event_display} shows the lego display of two dijet events, 
one with large ($R_{jj}=0.81$) and another with small ($R_{jj}=0.36$) dijet mass fractions.
Both event displays show some energy deposition well outside the jet regions, 
indicating that the events are not produced exclusively.

\vspace*{0.2cm}
In conclusion,
dijet exclusive production in DPE at the Tevatron can be used to set a baseline for the search of diffractive Higgs at the LHC. 
However, no exclusive dijet events have been found yet, and only a cross section upper limit was 
set at $\sigma_{excl}^{TeV}< 1.1 (0.03)$~nb for jets with $E_T>10 (25) ~{\rm GeV}$.

\section{Heavy Flavor Tagging}

\begin{figure}[h]
\epsfxsize=1.0\textwidth
\centerline{\epsfig{figure=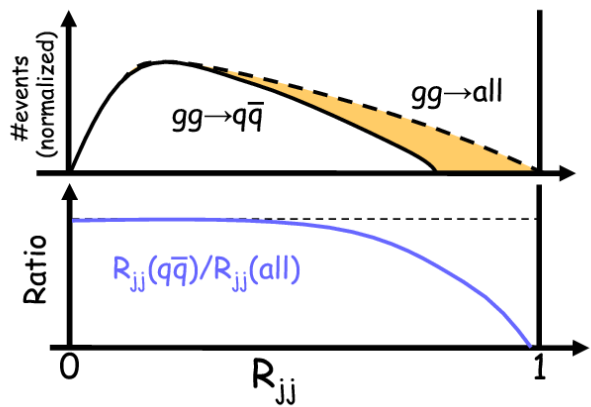,width=0.69\hsize}}
\caption{\label{fig:exclusivePlot}
\small
Schematic illustration of the behavior of dijet events
processes as a function of the mass fraction, $R_{jj}=M_{jj}/M_X$, 
i.e. the ratio of dijet mass divided by the invariant mass of the entire system.
Top: 
The spectrum of $q\bar{q}$ events (solid line) is suppressed in the high (exclusive) 
$R_{jj}$ region with respect to all dijet events (dashed line).
Bottom: the ratio of $q\bar{q}$ to all dijets is expected to fall in the exclusive 
region ($R_{jj}\rightarrow 1$) if exclusive dijet events are produced.}
\end{figure}

The quark/gluon composition of dijet final states can be used to provide additional hints on exclusive dijet production.
At leading order (LO), the exclusive $gg\rightarrow gg$ process is dominant, 
as the contribution from $gg\rightarrow q\bar{q}$ is strongly suppressed~\cite{khoze}.
In fact, the exclusive dijet cross section $\sigma_{excl} (gg\rightarrow q\bar{q})$ vanishes as $m_q/M^2\rightarrow 0$ ($J_z=0$ selection rule).
This condition is satisfied when quarks are light (such as $u$, $d$, or $s$ quarks), or when the dijet mass is much larger than the quark mass.
Thus, if the dijet mass is large enough compared to the $b$-quark mass, only gluon jets will be produced exclusively. 
This ``suppression'' mechanism can be used to extract an improved upper limit on the exclusive dijet cross section.
Figure~\ref{fig:exclusivePlot} illustrates the method that can be used to determine the heavy-flavor composition of the final sample.

\begin{figure}[h]
\epsfxsize=1.0\textwidth
\centerline{\epsfig{figure=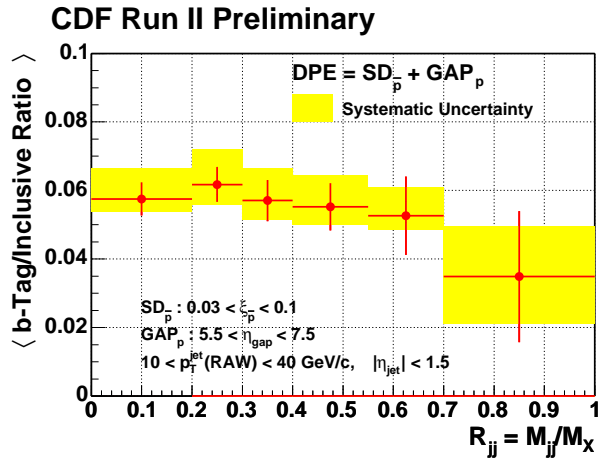,width=0.69\hsize}}
\caption{\label{fig:btagRatio}
\small
Ratio of b-tagged jets to all inclusive jets as a function of the mass fraction $R_{jj}$. The error band corresponds to the overall systematic uncertainty.
}
\end{figure}

The measured ratio, $D$, of $b$-tagged jets divided by all inclusive jet events is presented
in Figure~\ref{fig:btagRatio} as a function of the dijet mass fraction.
A decreasing trend is observed in the data in the large mass fraction region ($R_{jj}>0.7$), 
which may be an indication that the inclusive distribution contains an
exclusive dijet production component.
The fraction of $b$-tagged jets to inclusive jets,
$S=D^{>0.7}/D^{<0.4}$, is measured to be $0.59\pm0.33({\rm stat})\pm0.23({\rm syst})$.
Only four events are present in the last bin and a definite conclusion about exclusive production 
cannot be drawn due to the large statistical and systematical uncertainties.

\section{Conclusions}

Improved CDF forward detectors added new capabilities and extended our understanding of diffractive phenomena during 
Run~II at the Tevatron, in $p\bar{p}$ collisions at $\sqrt{s}=1.96$~TeV.
Exclusive production of dijet events has not yet been found in the data and stringent cross section limits have been set.
Further investigation of the quark/gluon composition of dijet final states has been exploited, and
a method of extracting the exclusive dijet production from inclusive data using bottom quark jets is presented.
The small number of events collected so far is not sufficient to identify exclusive dijet production.
More data to be collected at the Tevatron with a specifically designed trigger will hopefully help shed light on the exclusive process mechanism.

\end{document}